\documentclass[twocolumn,showpacs,aps,pra,superscriptaddresses]{revtex4}
\usepackage{graphicx,amsmath,amssymb}
\usepackage{dcolumn,fancyhdr}
\usepackage{bm,natbib}
\usepackage{times}
\usepackage{txfonts}
\usepackage{epstopdf, latexsym}

\usepackage{amsmath,epsfig}

\newcommand{\om}{\omega}

\newcommand{\g}{\gamma}

\begin{document}
\title{Distributing fully optomechanical quantum correlations}
\author{L. Mazzola$^{1,2}$ and M. Paternostro$^2$}
\affiliation{$^1$Turku Centre for Quantum Physics, Department of Physics and Astronomy, University of
Turku, FI-20014 Turun yliopisto, Finland\\
$^2$Centre for Theoretical Atomic, Molecular and Optical Physics, School of Mathematics and Physics, Queen's University Belfast, BT7 1NN Belfast, United Kingdom}

\begin{abstract}
We present a scheme to prepare quantum correlated states of two mechanical systems based on the {\it pouring} of pre-available all-optical entanglement into the state of two micro-mirrors belonging to remote and non-interacting optomechanical cavities. 
We show that, under realistic experimental conditions, 
the protocol allows for the preparation of a genuine quantum state of a composite mesoscopic system whose non-classical features extend beyond the occurrence of entanglement. We finally discuss a way to access such mechanical correlations.
\end{abstract}

\maketitle

\section{Introduction}
Although there is no fundamental constraint to the achievement of a quantum mechanical state of a macroscopic system, our daily experience is that we hardly observe any non-classical feature in intrinsically large or massive systems. A widely accepted justification for such difficulties relies on the paradigm of {\it decoherence}: it is not possible to perfectly isolate a system from its surrounding environment. This determines an uncontrollable flux of information between such parties that acts to the detriment of any non-classicality the system was endowed with. Such effect appears to be severe for large-scale systems. Nonetheless, the spectacular level of sophistication reached in the manipulation and control of complex systems has now enabled the exploration of quantum mechanical effects on an increasingly large scale in various physical contexts. Such endeavors have produced important results: from the preparation of moderate-amplitude optical Schr\"odinger cat states~\cite{GrangierNat07} to the engineering of multipartite entangled states in superconducting circuits~\cite{DiCarloNat09}, optically trapped atoms~\cite{BlochNat02} and Bose-Einstein condensates~\cite{OberthalerNat08}. 


Recently nano/micro-scale vibrating structures have received considerable attention as potential platforms for the realization of ultra-sensitive measurements and as test-beds for probing the foundations of quantum mechanics. Passive and feedback-based cooling of micromechanical devices~\cite{cooling}, strong optomechanical coupling~\cite{Gr?blacherNat09,TeufelNat11, Anetsberger10} and, to some extent, the preparation of the ground state of a micro-mechanical mode~\cite{ConnellNat09,TeufelArXiv, RocheleauNat10} have been demonstrated in some outstanding experiments. An optimistic approach to the progress of this area of investigation makes the preparation of highly non-classical mechanical states a foreseeable possibility. 
This has triggered proposals to prepare optomechanical and fully mechanical entangled states~\cite{EntScheme, Huang09, Braunstein03,VitaliPRL07,GirvinARXIV11}, coherent macroscopic superpositions~\cite{MarshallPRL03} and squeezed states of mechanical modes~\cite{SquezScheme} achieved with either a `gently' modulated driving or the use of non-classical light. A few steps towards hybridization through atomic-induced control of optomechanical systems have been performed~\cite{AtomControl}. 


In this paper we propose a scheme for the generation of a non-classical state of two spatially distant mechanical oscillators. The idea is to `distribute' non-classical correlations from a pair of electromagnetic fields to two remote and non-interacting mechanical modes. 
This is done by feeding two independent optomechanical cavities with a two-mode squeezed state of light~\cite{WallsMilburn}. We show how the light-matter interaction `transduces' the optical quantum correlations into fully mechanical ones, robustly against the dissipative mechanisms affecting the setup, the mechanical modes' masses and the operating temperature. Besides the fundamentally interesting result of achieving entanglement between two massive objects subjected to open-system dynamics, our proposal has also technological relevance. In fact, the scheme represents an experimentally viable strategy to distribute quantum correlations among spatially separated micro-mechanical devices. In this sense, our study embodies the continuous-variable counterpart of proposals for continuous-to-discrete-variable entanglement distribution~\cite{mauro}, realized through the (for this task) largely unexplored mechanism of radiation pressure coupling. Moreover, our work goes beyond the proposal put forward in Ref.~\cite{Braunstein03}, where an adiabatic Hamiltonian model was considered so as to `transfer' the whole optical state onto the mechanical one. We achieve non-classicality in a much less stringent interaction regime, which makes our flexible protocol different in principle and closer to the experimental reality.

The basic idea of our scheme suggests also a way to indirectly infer the non-classicality of the mechanical modes. In fact, we show that the distributed fully mechanical entanglement can be mapped out  onto the state of ancillary light modes and thus read through routine all-optical detection means. 
The feasibility of the proposal is verified using values for the relevant parameters in the problem borrowed from a recent experiment~\cite{Gr?blacherNat09}, which makes our proposal a foreseeable possibility for building up a table-top network of optomechanical nodes connected by shared quantum correlations.

The remainder of this paper is organized as follows. In Sec.~\ref{prot} we introduce the system at hand and describe our proposed protocol for entanglement distribution. Sec.~\ref{nonclas} is devoted to the quantification of the non-classical correlations in the purely mechanical system. Such an analysis will be performed by determining both entanglement and more general quantum correlations. Sec.~\ref{infer} described our proposal for entanglement inference, which is based on the use of a pair of ancillary optical modes. Finally, in Sec.~\ref{conc} we draw our conclusions. 

\section{The model and protocol}
\label{prot}

We consider two independent and non-interacting optomechanical systems, each consisting of a Fabry-Perot cavity composed of a heavy fixed input mirror and a lighter movable end one. The movable mirrors are modelled as single-mode quantum harmonic oscillators coupled via radiation-pressure to the corresponding intracavity field (see Fig.~1). Each cavity has length $L_i$ and frequency $\om_C^{i}$ and is pumped by a laser field with frequency $\om_L$. In addition, the two cavities are jointly pumped by a two-mode light field (each mode has frequency $\om_S$), prepared in an entangled state. The Hamiltonian of the two optomechanical devices, in a frame rotating at the frequency of the lasers, reads
\begin{equation}
\label{ham}
\begin{split}
\hat{H}{=}\!\!
\sum_{i=1,2}\!\!\hbar\delta^i\hat{n}_{i}{-}\hbar\chi_i \hat{n}_{i}\hat{q}_i{+}\!\left(\frac{\hat{p}^2_i}{2m_i}{+}\frac{m_i\om_{m}^{i 2}}{2} \hat{q}^2_{i}\right)\!{+} i \hbar \mathcal{E}_i(\hat{c}^{\dagger}_i{-}\hat{c}_i),
\end{split}\end{equation}
where $\hat{q}_{i}$ ($\hat{p}_{i}$) is the position (momentum) operator of the $i^\text{th}$ mirror, $\hat{c}_{i}$ ($\hat{c}^{\dagger}_{i}$) is the annihilation (creation) operator of the $i^\text{th}$ cavity field (whose photon-number operator and energy decay rate are $\hat{n}_{i}$ and $\kappa_i$, respectively), $\delta^i{=}\om_C^{i}{-}\om_L$ is the pump-cavity detuning. Moreover, $\omega_{m}^i$ and $m_i$ are frequency and mass of the $i^\text{th}$ mechanical oscillator and $\chi_i=\om_C^{i}/L_i$ is the coupling rate with the corresponding cavity field. Finally, we have $\mathcal{E}_i= \sqrt{\frac{2 \kappa_i\mathcal{P}_i}{\hbar\om_L^{i}}}$ with $\mathcal{P}_{1,2}$ the power of the pumping fields.

\begin{figure}
\includegraphics[scale=1.2]{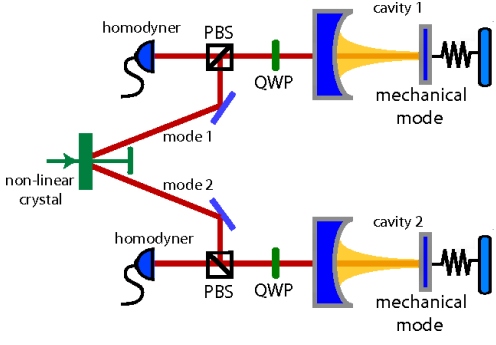} \caption{(Color online)
Sketch of the thought experiment: two optomechanical cavities are pumped by classical laser fields (not shown) and two-mode squeezed light, generated by feeding a nonlinear crystal. We show the symbols for polarization beam splitters (PBSs) and quarter wave-plates (QWPs) that are used to direct the polarized input squeezed light (output field) to the cavity (the homodyne-detection stage).}
\label{scheme}
\end{figure}


The system is affected by cavity losses and decoherence induced by the thermal Brownian motion of the mechanical oscillators, which are damped at a rate $\gamma_{m}^i$. The model embodied by Eq.~(\ref{ham}) encompasses a dynamics of non-trivial solution due to the explicitly nonlinear nature of the radiation-pressure term. A considerable simplification, though, comes from having intense fields feeding the cavities, thus allowing us to linearize the fields' and mirrors' operators around their respective steady-state. 
 The equations of motion can then be cast into the compact form~\cite{WallsMilburn}
\begin{equation}\label{Langevin}
\partial_t \hat{\mathbf{f}}_i= \mathbf{K}_i\hat{\mathbf{f}}_i+\hat{\mathbf{n}}_i,~~~~~(i=1,2)
\end{equation}
where $\hat{\mathbf{f}}_i^T{=}(\delta \hat{Q}_i,\delta \hat{P}_i,\delta \hat{x}_i,\delta \hat{y}_i)$ 
is the ordered vector of the fluctuations of the dimensionless quadrature operators $\hat{Q}_i=\hat{q}_i\sqrt{m_i\om_{m}^{i}/\hbar}$ and $\hat{P}_i=\hat{p}_i/\sqrt{\hbar m_i \om_{m}^i}$ for mechanical mode $i$ and $\delta \hat{x}_i=(\delta \hat{c}_i^{\dagger}+\delta \hat{c}_i)/\sqrt{2}$, $\delta \hat{y}_i=i(\delta \hat{c}_i^{\dagger}-\delta \hat{c}_i)/\sqrt{2}$ for the corresponding cavity fields. Each $4\times4$ kernel matrix $\mathbf{K}_i$ reads 
\begin{equation}
\label{Ai}
\mathbf{K}_i{=}\!\!\left[\begin{array}{cccc}
\!\!\!0 & \om_{m}^i & 0 & 0\!\!\!\\
\!\!\!-\om_{m}^i & -\gamma_{m}^i & 2g_i\Re[c^i_s] & -2 g_i \Im[c_{s}^{i}]\!\!\!\\
\!\!\!-2g_i \Im[c_{s}^{i}] & 0 & -\kappa_i & \Delta_i\!\!\!\\
\!\!\!-2i g_i\Re[c_{s}^{i}] & 0 & -\Delta_i & -\kappa_i\!\!\!
\end{array}\right]~~(i=1,2)
\end{equation}
with $g_i{=}\chi_i\sqrt{\hbar/(2m_i\om_{m}^i)}$ being an effective coupling rate, $c_{s}^{i}{=}{\mathcal{E}_i}/({\kappa_i+i\Delta_i})$ the amplitude of cavity field $i$ and $\Delta_i{=}\om_C^{i}{-}\om_L{-}\chi q_{s}^i$ the corresponding effective cavity-laser detuning. Finally $q_{s}^i{=}\frac{\hbar\chi_i |c_{s}^{i}|^2}{m_i\om^{i2}_{m}}$ is the steady state displacement of the mechanical mode $i$. The last term in Eq.~(\ref{Langevin}) is the vector of input noise $\mathbf{n}^T_j{=}(0,\hat{\xi}_i,\sqrt{2\kappa_i}\delta\hat{x}_{in}^i,\sqrt{2\kappa_i}\delta\hat{y}_{in}^i)$, where $\hat{\xi}_i$ is the zero-mean Langevin force operator accounting for the Brownian motion affecting the mechanical mode $i$. 
For large mechanical quality factors, such term is correlated as $\langle\hat{\xi}_i(t)\hat{\xi}_i(t')\rangle{=}(2\g_m^i k_B T_i) \delta(t{-}t')/\om_m^i$, with $k_B$ the Boltzmann constant and $T_i$ the temperature of the $i^\text{th}$ mechanical bath, while $\delta \hat{x}_{in}^i{=}(\delta \hat{c}_{in}^{i\dagger}{+}\delta \hat{c}_{in}^i)/\sqrt{2}$ and $\delta \hat{y}_{in}^i{=}i(\delta \hat{c}_{in}^{i\dagger}-\delta \hat{c}_{in}^i)/\sqrt{2}$ are the quadratures of the input noise to a cavity.

Clearly, the dynamics encompassed by Eq.~(\ref{Langevin}) would give rise to two independent evolutions, one for each optomechanical device. This seemingly prevents the establishment of any quantum correlation between the two mechanical modes, unless a pre-available resource is used. In fact, as proven in Refs.~\cite{mauro,AdessoPRL10}, the sharing of an off-line prepared entangled resource can be used in order to set entanglement between remote and non-interacting subsystems, subjected to bilocal interactions with the components of such entangled aid. We thus consider the two cavities pumped by non-classically correlated light as embodied by a two-mode squeezed vacuum state~\cite{commentoAga}. Each mode of such pump has frequency $\om_S{=}\om_L{+}\om_m$ (where we take equal mechanical frequencies $\om_m=\om_m^{1,2}$) so that the corresponding input-noise correlations ${\bf C}_j=(\langle\delta\hat{c}^{j\dagger}_{in}\delta\hat{c}^j_{in}\rangle,\langle\delta\hat{c}^j_{in}\delta\hat{c}^{j\dagger}_{in}\rangle,\langle\delta\hat{c}^j_{in}\delta\hat{c}^k_{in}\rangle,\langle\delta\hat{c}^{j\dagger}_{in}\delta\hat{c}^{k\dagger}_{in}\rangle)$ [for $j{\neq}k{=}1,2$] can be expressed as
${\bf C}_j={\bf R}(t,t')\delta(t-t')$
%
with ${\bf R}(t,t'){=}(N,N+1,e^{-i \om_m (t-t')} M, e^{i \om_m (t-t')} M^*)$ a vector determined by the parameters $N{=}\sinh^2{r}$ and $M{=}\sinh{r}\cosh{r}e^{i\phi}$ that characterize the squeezed state ($r$ and $\phi$ are the modulus and phase of the squeezing parameter). From now on, without affecting the generality of our study, we take $\phi=0$.

\begin{figure}
\includegraphics[scale=0.55]{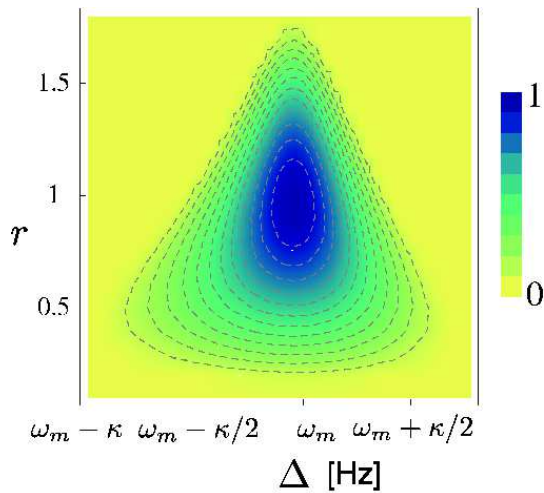} \caption{(Color online) Logaritmic negativity against the effective detuning $\Delta$ and the dimensionless squeezing parameter $r$ for $T=2\ {\rm mK}$. The remaining set of parameters is borrowed from the recent experiment reported in Ref. \cite{Gr?blacherNat09}.}
\label{DensityPlot1}
\end{figure}
Although the analytic expressions of the elements of $\hat{\bf f}_i$ can be found straightforwardly, they are too lengthy to be reported here. However, they can be handly used so as to calculate the correlation functions of the relevant operators of the system as
\begin{equation}
\label{cv}
V_{\alpha\beta}(t)=\frac{1}{4 \pi^2}\int d\om \int d\Omega e^{-i(\om+\Omega)t} V_{\alpha\beta}(\om,\Omega)
\end{equation}
with $V_{\alpha\beta}(\om,\Omega)=\langle\{\hat{g}_{\alpha}(\om),\hat{g}_{\beta}(\Omega)\}\rangle/2$ ($\alpha,\beta{=}1,..,8$) the frequency-domain correlation function between elements $\alpha$ and $\beta$ of $\hat{\bf g}{=}(\delta \hat{Q}_1,\delta \hat{P}_1,\delta \hat{Q}_2,\delta \hat{P}_2,\delta \hat{x}_1,\delta \hat{y}_1,\delta \hat{x}_2,\delta \hat{y}_2)$. The matrix ${\bf V}(t)$ with elements defined as in Eq.~\eqref{cv} embodies the time-dependent covariance matrix of the whole system. 
This is an important remark as the enforced linearity of the model treated here guarantees that the dynamical map arising from the solution of Eq.~\eqref{Langevin} preserves the Gaussian nature of any input state. We consider a Gaussian input noise (the two-mode squeezed vacuum state) and a thermal state of the mechanical modes determined by the quantum Brownian motion. As Gaussian states are fully determined by their associated covariance matrix, the convenience of such description is apparent. 

\section{Non-classical correlations}
\label{nonclas}

We are now in a position to assess the non-classical correlations settled between the two mechanical modes. We do this using a twofold approach: first we quantify the entanglement shared by mechanical modes $1$ and $2$, demonstrating the effectiveness of the process put forward here. Second, we demonstrate that the parameter-domain where purely mechanical non-classical correlations can be observed extends far beyond the one where entanglement is non-zero. Let us start studying the degree of entanglement between the mechanical modes. Clearly, while the achievement of a non-zero degree of entanglement would show the plausible nature of our proposal as a way to distribute mechanical entanglement across a network of optomechanical nodes, its effectiveness will be evaluated in terms of the ratio between the set mechanical entanglement and the amount {\it consumed} as a resource. We use the logarithmic negativity as a measure of entanglement in the state of our system. Given the covariance matrix ${\bm V}_m$ of the mechanical system (obtained from ${\bf V}$ by taking the first four rows and columns), the logarithmic negativity is found as $E{=}\max[0,-\ln2(\nu_-)]$ with $\nu_{-}$ the smallest eigenvalue of the symplectic spectrum of $\mathbf{V}^P_m=\mathbf{P}\mathbf{V}_m\mathbf{P}$ with ${\bf P}{=}\openone{\oplus}\mathbf{\sigma}_z$  ($\mathbf{\sigma}_i$ is the $i=x,y,z$ Pauli matrix)~\cite{Symplectic}. 

\begin{figure}
\includegraphics[scale=0.3]{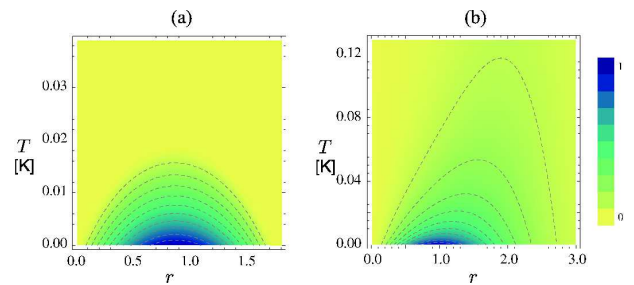} \caption{(Color online)
{\bf (a)} Logaritmic negativity against the dimensionless squeezing parameter $r$ and the temperature of the mechanical bath $T$ (in Kelvin) for $\Delta=\om_m$. {\bf (b)} Gaussian quantum discord against squeezing and temperature for $\Delta=\om_m$. Non-classicality of the mechanical modes is found up to $T=0.12 {\rm \ K}$.}
\label{InOut}
\end{figure}

As we aim at proposing a viable strategy for the settlement of all-mechanical entanglement, it is important to test our protocol against current experimental capabilities. Therefore, in our simulations we have used parameters taken directly from a very recent experiment~\cite{Gr?blacherNat09}. In particular, we have taken identical mechanical modes with $\om_m/2\pi=947$~KHz, $m_{1,2}=145{\rm\ ng}$ and $\gamma^{1,2}_m/2\pi=140$~Hz and cavities having length $L_{1,2}=25$~{mm}, wavelength $1064$~nm, decay rate $\kappa/2\pi=215$~KHz and pumped by laser fields of $11$~mW power~\cite{stab}. The free variables in our simulations are the experimentally tunable detuning $\Delta_{1,2}{=}\Delta$, the squeezing parameter $r$ and the initial temperature of the mechanical mirrors. Fig.~\ref{DensityPlot1} shows the logarithmic negativity for $T=2\ \mathrm{mK}$ and reveals that the mechanical entanglement is maximum close to the frequency $\om_m$ and non-zero in a region of $\Delta$ that is wide roughly $\kappa$. A non-trivial dependence on $r$ is also observed. As the squeezing increases, the region of values of $\Delta$ where $E\neq{0}$ narrows. At the same time, $E$ first slowly increases with $r$, reaching a maximum at $r\simeq{1}$ and quickly decreasing for larger values ($E$ disappears completely for $r\gtrsim1.6$). This behavior reminds of the analogous features observed in the transfer of entanglement to two remote qubits~\cite{mauro}. However, while in that case the responsible for such features was the mismatch in the Hilbert-space dimensions, here the explanation is that with increasing squeezing also the thermal noise entering each cavity is enhanced~\cite{sq}. We emphasize that two-mode squeezed states with squeezing parameters of the order of unity are well within experimental reach, as demonstrated, for example, in Ref.~\cite{LauratPRA05}.

\begin{figure}
\includegraphics[scale=0.8]{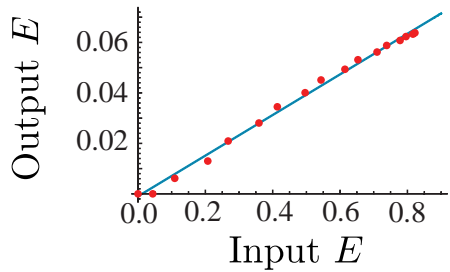} \caption{(Color online) Entanglement input-output relation showing the mapping of mechanical entanglement onto optical one shared by the auxiliary modes. The quantities on the horizontal and vertical axes are dimensionless.}
\label{DensityPlot2}
\end{figure}

We now investigate the behaviour of mechanical non-classical correlations beyond entanglement, as assessed by Gaussian quantum discord~\cite{AdessoDattaPRL10}.
Quantum discord is defined from the discrepancy that two classically equivalent expressions of mutual information attain when extended to the quantum domain \cite{Zurek}. It is connected to the information on a system that cannot be retrieved by performing measurements only on a part of it. For Gaussian continuous-variable systems, the evaluation of discord relies on an optimization procedure over the set of any possible single-mode Gaussian measurements. For a two-mode Gaussian state having covariance matrix $\boldsymbol\sigma{=}\left[\begin{matrix}\boldsymbol{\alpha}_1 &\boldsymbol{\gamma}\\  \boldsymbol{\gamma}^T & \boldsymbol{\alpha}_2\end{matrix}\right]$, where $\boldsymbol\alpha_1$ ($\boldsymbol\alpha_2$) and $\boldsymbol\gamma$ are $2\times2$ matrices accounting for the local variances of mode $1$ ($2$) and the inter-mode correlations, the Gaussian quantum discord is $\mathcal{D}=f(\sqrt{A_2})-f(\mu_-)-f(\mu_+)+\inf_{\boldsymbol{\sigma}_0} f(\sqrt{{\rm det}\boldsymbol\epsilon})$. Here, $f(x)=(\frac{x+1}{2})\log[\frac{x+1}{2}]-(\frac{x-1}{2})\log[\frac{x-1}{2}]$, $A_2=\det\boldsymbol\alpha_2$, $\mu_{\pm}$ are the symplectic eigenvalues of $\boldsymbol\sigma$, $\boldsymbol{\epsilon}=\boldsymbol{\alpha_1}-\boldsymbol{\gamma}(\boldsymbol{\alpha_2}+\boldsymbol\sigma_0)^{-1} \boldsymbol\gamma^T$ is the Schur complement of ${\bm\alpha}_1$, and $\boldsymbol\sigma_0$ is a single-mode rotated squeezed state.
Fig.~\ref{DensityPlot2} compares the behavior of $E$ (part ({\bf a})) and ${\cal D}$ (part ({\bf b})) for the two mechanical modes when studied against $T$ and $r$ for $\Delta=\omega_m$. The two indicators of non-classicality reveal rather different trends when analyzed against the squeezing parameter. At low temperatures, the maximum of ${\cal D}$ is attained for $r\simeq1$, while for $T{\gtrsim}10 {\rm \ mK}$ (when entanglement disappears), ${\cal D}$ grows with the squeezing. Remarkably non-classical correlations still persists and are significantly non zero for temperature higher than $0.1$K, thus demonstrating that the model addressed in our proposal is able to set robust and general non-classical correlations in quite an ample region of the working parameters. 


\section{Entanglement inference}
\label{infer}

Here we put forward a proposal to reveal the entanglement set between the mechanical modes. Our scheme relies on the use of two auxiliary light modes, each prepared in a coherent state and interacting with the respective optomechanical cavity. The mechanical modes are assumed to be prepared in an entangled state of the sort addressed above. The concept at the basis of our inference scheme is similar to the one underlying the entanglement-setting one: the optomechanical interaction acts as a transducer, this time changing entanglement from mechanical to fully optical. As the ancillary modes are initially uncorrelated and do not interact directly, the creation of optical entanglement can only be a signature of the pre-existence of mechanical one. Differently from~\cite{VitaliPRL07}, our scheme does not require an adiabatic light-mirror interaction nor any weak coupling approximation between mechanical and photonic ancillary modes.

We take a fully dynamical approach based on the solution of the equation satisfied by the covariance matrix $\tilde{\mathbf{V}}$ of the mechanical modes and the auxiliary fields, which can be shown to read  $\partial_t \tilde{\mathbf{V}}=\mathbf{L}\tilde{\mathbf{V}}+\tilde{\mathbf{V}}\mathbf{L}^T+\mathbf{D}$. Here $\mathbf{L}$ is the kernel of the Langevin equations of each ancilla-mirror system ($\mathbf{L}{=}\oplus^2_{i{=}1}\mathbf{K}_i$ if we take the same parameters of the intracavity modes for the auxiliary ones), and $\mathbf{D}={\rm Diag}[0,\g_m(2 \bar{n}+1),\kappa,\kappa,0,\g_m(2 \bar{n}+1),\kappa,\kappa]$  is the noise matrix with $\bar{n}=(\exp\{\hbar\om_m/k_B T\}-1)^{-1}$. 
Figure~\ref{InOut} shows the conversion of input mechanical entanglement in output optical one at $\Delta=\omega_m^{1,2}$, $T=2$~mK and $r\in[0,2]$. Remarkably, the output optical entanglement turns out to be linearly dependent on the input mechanical one. As the former can be determined by means of homodyne measurements on the ancillary modes, this suggests the possibility to quantify mechanical entanglement by subjecting the reconstructed all-optical one to a post-processing procedure that rescales it according to the linear behavior highlighted above.


\section{Conclusions}
\label{conc}

We have introduced a protocol for the engineering of spatially distributed mechanical non-classicality through a simple bi-local radiation-pressure mechanism. By `consuming' a pre-available resource of entanglement, we have achieved a highly non-classical state of two mechanical oscillators, suitable for the distribution of entanglement and general quantum correlation in a network of continuous-variable systems. We have examined both the stationary and fully time-dependent dynamics of the system, finding robustness of the enforced quantumness in a wide range of operating parameters. We have also discussed a rather promising scheme for the inference of the distributed fully mechanical quantum correlations based on a simple all-optical detection scheme and straightforward post-processing of data. Although only hinted in our work, it will be very interesting to explore deeply the potential of our scheme as a mean to merge optomechanics and the theory of quantum memories~\cite{KimbleNat08} towards the construction of a quantum network whose nodes are embodied by high-quality mechanical system~\cite{PainterNat09,Lukin}.

\acknowledgments

We acknowledge G. Adesso, J. Eisert, S. Maniscalco and A. Mari for useful discussions. LM is grateful to the School of Engineering \& Physical Sciences, Heriot-Watt University, for kind hospitality during the completion of this work. This work was supported by the Magnus
Ehrnrooth Foundation, the UK EPSRC (EP/G004759/1) and the British Council.



\begin{thebibliography}{99}
\bibitem{commentoAga} In Ref.~\cite{Huang09} a single-mode squeezed state was used to entangle two mirrors belonging to the same ring cavity.
\bibitem{Symplectic}
The symplectic spectrum of $\mathbf{V}^P$ is given by the eigenvalues of the matrix $|i(\oplus^2_{i{=}1}\mathbf{\sigma}^i_y)\mathbf{V}^P|$.
\bibitem{sq} The reduced state of a two-mode squeezed vacuum  is a thermal state having an average number of photons proportional to the squeezing parameter.
\end{thebibliography}
\end{document}